\documentclass{iau}
\usepackage{graphicx}
\usepackage{natbib}
\usepackage{amsmath, amssymb} 
\usepackage{bm}       
\usepackage{mathtools} 
\usepackage{empheq} 
\usepackage{amsfonts}
\usepackage{mathrsfs} 

\usepackage[usenames]{xcolor}

\usepackage[normalem]{ulem}
\DeclareMathAlphabet{\mathpzc}{OT1}{pzc}{m}{it}

\newcommand{\dif}[2]{\frac{\mathrm{d} #1}{\mathrm{d} #2}}

\newcommand{\prior}{\mathscr{P}}
\newcommand{\likeli}{\mathscr{L}}

\newcommand{\hamHMC}{\mathscr{H}}
\newcommand{\potHMC}{\mathscr{E}}
\newcommand{\kinHMC}{\mathscr{K}}
\newcommand{\masHMC}{\mathscr{M}}
\newcommand{\momHMC}{\mathpzc{p}}
\newcommand{\posHMC}{\mathpzc{q}}\newcommand{\bs}[1]{{\bm{#1}}}
\newcommand{\bx}{\bs{x}}   
\newcommand{\bq}{\bs{q}}   
\newcommand{\bv}{\bs{v}}   
\newcommand{\hatx}{\hat{\bs{x}}} 
\newcommand{\obs}{\mathrm{obs}}
\newcommand{\displace}{\bs{s}}
\newcommand{\mpc}{\mathrm{Mpc}}
\newcommand{\mpch}{~h^{-1} \mpc}

\newcommand{\barcode}{\textsc{barcode}}
\newcommand{\zspace}{\bs{\zeta}}
\newcommand{\displacez}{{\bs{s}_\zeta}}

\title[Bayesian Cosmic Web Reconstruction] 
{Bayesian Cosmic Web Reconstruction:\\ {\Large BARCODE for Clusters}}

\author[E.G.P. Bos, R. van de Weygaert, F.S. Kitaura \& M. Cautun]   
{E. G. Patrick Bos$^1$,
 Rien van de Weygaert$^1$,
 Francisco Kitaura$^2$,
 \and
 Marius Cautun$^3$}

\affiliation{
$^1$Kapteyn Astron. Inst., Univ.\ of Groningen,
Groningen, the Netherlands, email: {\tt pbos@astro.rug.nl} \\[\affilskip]
$^2$Leibniz-Inst.\ f\"ur Astrophysik Potsdam (AIP), An der Sternwarte 16,
14482 Potsdam, DE \\[\affilskip]
$^3$Inst.\ for Computational Cosmology, Univ.\ of Durham, 
Durham DH1 3LE, United Kingdom
}
\pubyear{2014}
\volume{308}  
\pagerange{0--0}
\jname{the Zeldovich Universe: Genesis and Growth of the Cosmic Web}
\editors{R. van de Weygaert, S.F. Shandarin, E. Saar \& J. Einasto, eds.}

\begin{document}

\maketitle

\begin{abstract}
We describe the Bayesian \barcode\ formalism that has been designed towards the reconstruction of 
the Cosmic Web in a given volume on the basis of the sampled galaxy cluster distribution. Based on the 
realization that the massive compact clusters are responsible for the major share of the large 
scale tidal force field shaping the anisotropic and in particular filamentary features in the 
Cosmic Web. Given the nonlinearity of the constraints imposed by the cluster configurations, we 
resort to a state-of-the-art constrained reconstruction technique to find a proper statistically 
sampled realization of the original initial density and velocity field in the same cosmic region. 
Ultimately, the subsequent gravitational evolution of these initial conditions towards the 
implied Cosmic Web configuration can be followed on the basis of a proper analytical 
model or an $N$-body computer simulation. The \barcode\ formalism includes an implicit treatment for redshift 
space distortions. This  enables a direct reconstruction on the basis of observational data, without 
the need for a correction of redshift space artifacts. In this contribution we provide a general overview 
of the the Cosmic Web connection with clusters and a description of the Bayesian \barcode\ formalism. We 
conclude with a presentation of its successful workings with respect to test runs based on a simulated 
large scale matter distribution, in physical space as well as in redshift space.

\end{abstract}

\firstsection

\section{Introduction}
On Megaparsec scales, matter and galaxies have aggregated into a complex network of interconnected filaments and walls. 
This network, which has become known as the \emph{Cosmic Web} \citep{bondweb96}, contains structures from a few megaparsecs up 
to tens and even hundreds of megaparsecs of size. It has organized galaxies and mass into a wispy web-like spatial arrangement, 
marked by highly elongated filamentary, flattened planar structures and dense compact clusters surrounding large near-empty void 
regions. Its appearance has been most dramatically illustrated by the maps of the nearby cosmos produced by large galaxy redshift 
surveys such as the 2dFGRS, the SDSS, and the 2MASS redshift surveys \citep{Col03,TSB04,HMM12}, as well as by recently produced 
maps of the galaxy distribution at larger cosmic depths such as VIPERS \citep{guzzo2013,guzzo2014}.

The Cosmic Web is one of the most striking examples of complex geometric patterns found in nature, and certainly the 
largest in terms of size. Computer simulations suggest that the observed cellular patterns are a prominent and natural aspect of 
cosmic structure formation through gravitational instability \citep{Pee80}. According to the \emph{gravitational instability scenario}, 
cosmic structure grows from tiny primordial density and velocity perturbations. Once the gravitational clustering process has progressed 
beyond the initial linear growth phase, we see the emergence of complex patterns and structures in the density field. 
Within this context, the Cosmic Web is the most prominent manifestation of the Megaparsec scale tidal force field. 

The recognition of the \emph{Cosmic Web} as a key aspect in the emergence of structure in the Universe came with early analytical 
studies and approximations concerning the emergence of structure out of a nearly featureless primordial Universe. 
In this respect the Zel'dovich formalism~\citep{zeldovich70} played a seminal role. It led to view of 
structure formation in which planar pancakes form first, draining into filaments which in turn drain into 
clusters, with the entirety forming a cellular network of sheets. 

As borne out by a large sequence of N-body computer experiments of cosmic structure formation, web-like patterns in the overall 
cosmic matter distribution do represent a universal but possibly transient phase in the gravitationally driven emergence and 
evolution of cosmic structure. N-body calculations have shown that web-like patterns defined by prominent anisotropic filamentary 
and planar features --- and with characteristic large underdense void regions --- are a natural manifestation of the gravitational 
cosmic structure formation process. Within the context of gravitationally driven structure formation, the formation and 
evolution of anisotropic structures is a direct manifestation of the gravitational tidal forces induced by the 
inhomogeneous mass distribution. It is the anisotropy of the force field and the resulting deformation of the matter 
distribution which which are at the heart of the emergence of the web-like structure of the mildly nonlinear 
mass distribution. 

Interestingly, for a considerable amount of time the emphasis on anisotropic collapse as agent for forming and 
shaping structure was mainly confined the Soviet view of structure formation, Zel'dovich's pancake picture, and 
was seen as the rival view to the hierarchical clustering picture which dominated the western view. The successful 
synthesis of both elements culminated in the Cosmic Web theory \citep{bondweb96}, which stresses the dominance of 
filamentary shaped features instead of the dominance of planar pancakes in the pure Zel'dovich theory. 

Perhaps even more important is the identification of the intimate dynamical relationship between the filamentary patterns 
and the compact dense clusters that stand out as the nodes within the cosmic matter distribution: filaments as 
cluster-cluster bridges \citep[also see][]{weyedb96,bondweb1998,colberg05,weybond2008}. In the overall cosmic 
mass distribution, clusters --- and the density peaks in the primordial density field that are their precursors --- stand 
out as the dominant features for determining and outlining the anisotropic force field that generates the cosmic 
web. 

In this study, we pursue this observation and investigate in how far we should be able to outline in a given 
volume of the observed Universe the web-like dark matter distribution on the basis of the observed cluster distribution. 
The challenge of this question is that of translating a highly biased population of compact massive clusters, the 
result of a highly nonlinear evolution, into the corresponding implied mildly nonlinear Cosmic Web. This inversion 
problem calls in several state-of-the-art statistical inversion techniques, in combination with advanced analytical 
prescriptions for the mildly nonlinear gravitational evolution of the cosmic density field. Before describing the 
\barcode\ Bayesian reconstruction formalism, we will first provide the context of the gravitationally induced formation 
of filaments and walls. 

\subsection{Tidal Dynamics of the Cosmic Web}
Within the context of gravitational instability, it are the gravitational tidal forces that establish the relationship between some of the most 
prominent manifestations of the structure formation process. When describing the dynamical evolution of a region in the density field it is useful 
to distinguish between large scale ``background'' fluctuations $\delta_{\rm b}$ and small-scale fluctuations $\delta_{\rm f}$. Here, we are 
primarily interested in the influence of the smooth large-scale field.  

To a good approximation the smoother background gravitational force ${\bf g}_{\rm b}({\bf x})$ in and around the mass element includes three components. The \emph{bulk force} ${\bf g}_{\rm b}({\bf x}_{pk})$ is responsible for the acceleration of the mass element as a whole. Its divergence 
($\nabla \cdot {\bf g}_{\rm b}$) encapsulates the collapse of the overdensity while the tidal tensor $T_{ij}$ quantifies its deformation,  
\begin{equation}
g_{\rm b,i}({\bf x})\,=\,g_{\rm b,i}({\bf x}_{pk})\,+\,a\,\sum_{j=1}^3\,\left\{{\displaystyle 1 \over \displaystyle 3 a}(\nabla \cdot {\bf g}_{\rm b})({\bf x}_{pk})\,\delta_{\rm ij}\,-\,T_{ij}\right\} (x_j - x_{pk, {\rm j}})\,.
\end{equation}
The tidal shear force acting over the mass element is represented by the (traceless) tidal tensor $T_{ij}$,
\begin{eqnarray}
T_{\rm ij}&\,\equiv\,&-{\displaystyle 1 \over \displaystyle 2 a}\, 
\left\{ {\displaystyle \partial g_{{\rm b},i} \over \displaystyle \partial x_i} +
{\displaystyle \partial g_{{\rm b},j} \over \displaystyle \partial x_j}\right\}\,+\,
{\displaystyle 1 \over \displaystyle 3 a} (\nabla \cdot {\bf g}_{\rm b}) 
\,\delta_{\rm ij}
\end{eqnarray}
in which the trace of the collapsing mass element, proportional to its overdensity $\delta$, dictates its contraction (or expansion). For a cosmological 
matter distribution the close connection between the local force field and global matter distribution follows from the expression of the tidal tensor 
in terms of the generating cosmic matter density fluctuation distribution $\delta({\bf r})$ \citep{weyedb96}:
\begin{eqnarray}
&& T_{ij}({\bf r})={\displaystyle 3 \Omega H^2 \over \displaystyle 8\pi}\,
\int {\rm d}{\bf r}'\,\delta({\bf r}')\ \left\{{\displaystyle 3 (r_i'-r_i)(r_j'-r_j) -
|{\bf r}'-{\bf r}|^2\ \delta_{ij} \over \displaystyle |{\bf r}'-{\bf r}|^5}\right\} - {\frac{1}{2}}\Omega H^2\ \delta({\bf r},t)\ \delta_{ij} . \nonumber 
\label{eq:quadtide}
\end{eqnarray}
The Megaparsec scale tidal shear forces are the main agent for the contraction of matter into the sheets and 
filaments which trace out the Cosmic Web. The anisotropic contraction of patches of matter depends sensitively 
on the signature of the tidal shear tensor eigenvalues. With two positive eigenvalues and one negative,  
$(-++)$, we will see strong collapse along two directions. Dependent on the overall overdensity, along the 
third axis collapse will be slow or not take place at all. Likewise, a sheetlike membrane will be the product of 
a $(--+)$ signature, while a $(+++)$ signature inescapably leads to the full collapse of a density peak into 
a dense cluster. 

\begin{figure*}[b]
\begin{center}
 \includegraphics[width=13.0cm]{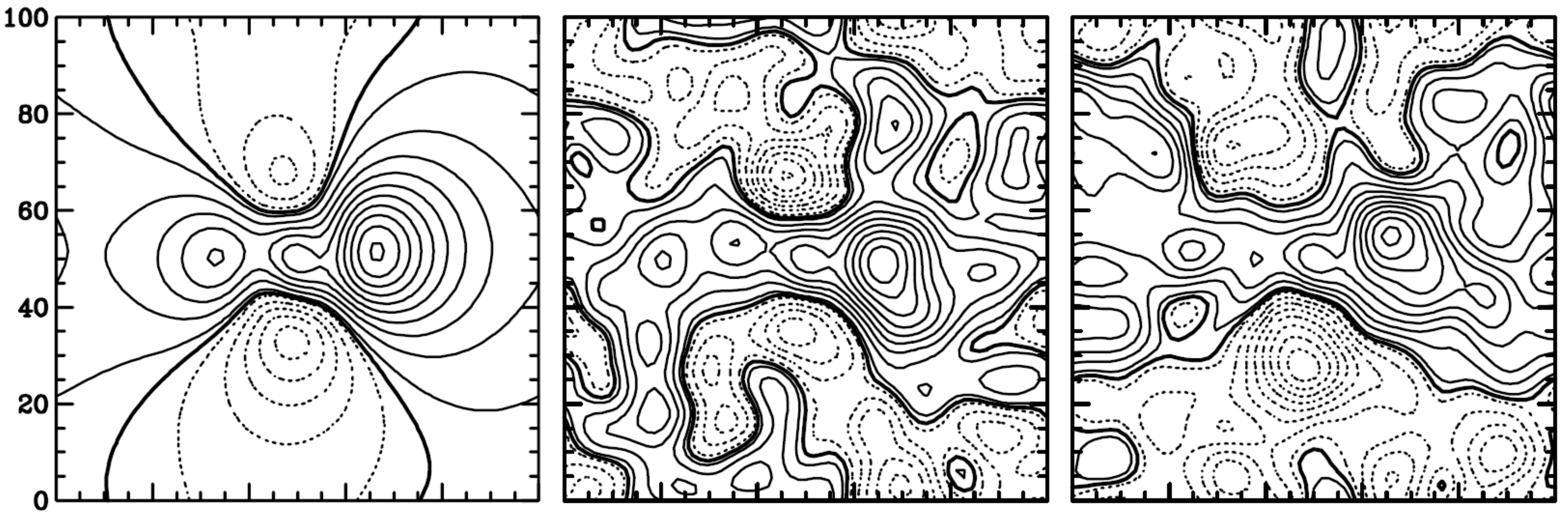} 
 \caption{The implied configuration in a Gaussian random field around a location with 
filamentary tidal compression: clearly visible are two protocluster density peaks that 
generate the specified tidal force field. This is the typical configuration out of which 
a filamentary bridge between two massive clusters will evolve. 
From: van de Weygaert \& Bertschinger 1996}
\label{fig:tidalshear_peaks}
\end{center}
\end{figure*}

\section{Clusters and the Megaparsec Tidal Force Field}
Given our understanding of the tidal force configuration that will induce collapse into a filamentary configuration, 
we may deduce what a typical mass distribution configuration generating such a force field. Figure~\ref{fig:tidalshear_peaks} 
shows that in the primordial Gaussian density field a generic configuration generating such a quadrupolar force field is 
that of two prominent density peaks \citep{weyedb96}. In the subsequent nonlinear evolution, these evolve into two massive clusters. 
Meanwhile, the mass in the intermediate region is gravitationally contracted towards an elongated filamentary configuration. 

\begin{figure*}[b]
\begin{center}
\vskip 0.25truecm
\mbox{\hskip -0.0truecm\includegraphics[width=12.1cm]{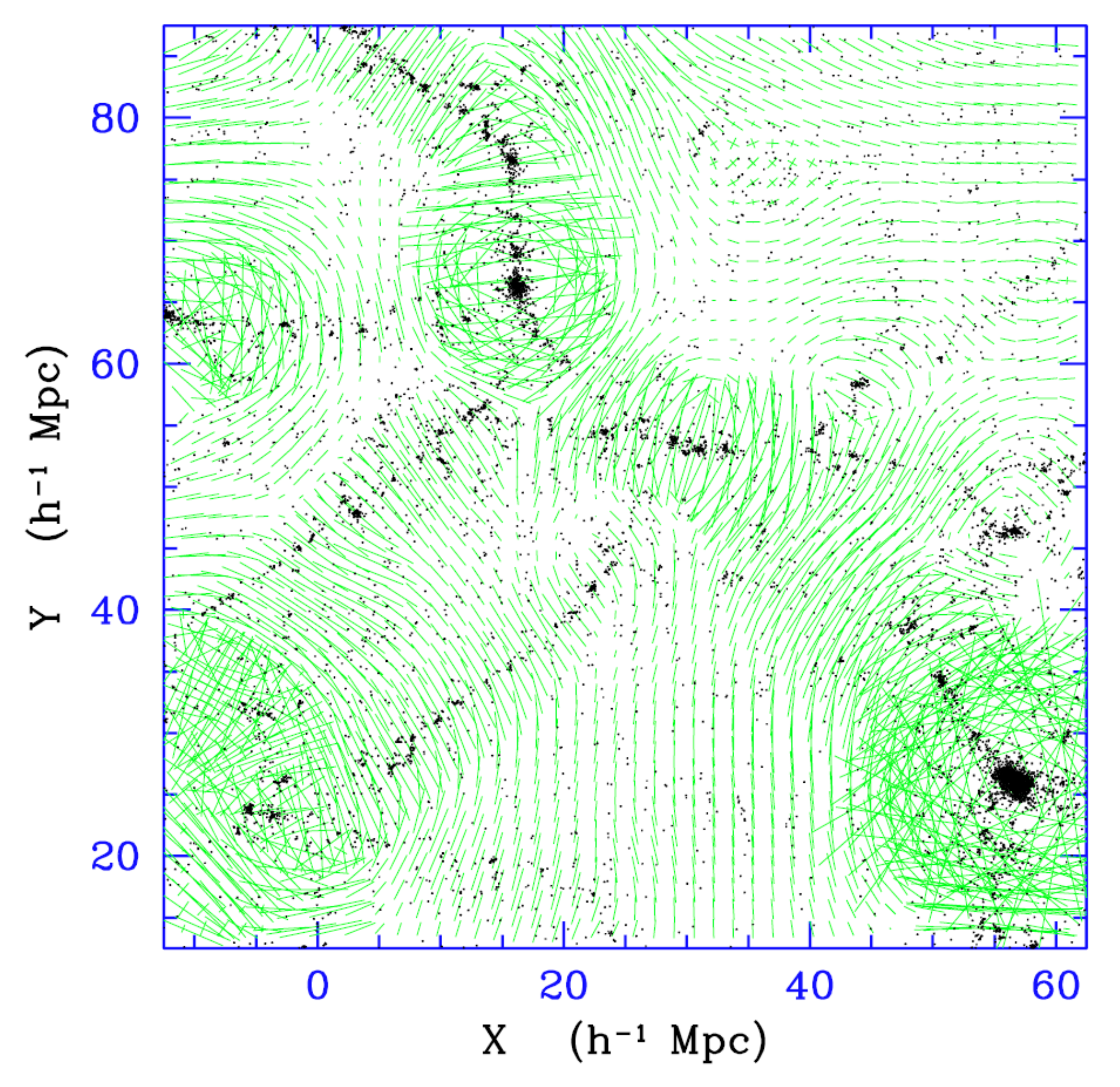}}
\vskip -0.5truecm
\caption{The relation between the \emph{Cosmic Web}, the clusters at the nodes in this network and 
the corresponding compressional tidal field pattern. It shows the matter distribution at the 
present cosmic epoch, along with the (compressional component) tidal field bars in a slice through 
a simulation box containing a realization of cosmic structure formed in an open, $\Omega_{\circ}=0.3$, Universe 
for a CDM structure formation scenario (scale: $R_G=2h^{-1}\hbox{\rm Mpc}$). The frame shows structure in a 
$5h^{-1}\hbox{\rm Mpc}$ thin central slice, on which the related tidal bar configuration is superimposed. 
The matter distribution, displaying a pronounced web-like geometry, is clearly intimately linked with a 
characteristic coherent compressional tidal bar pattern. From: van de Weygaert 2002} 
\label{fig:cosmicwebtide}
\end{center}
\end{figure*}

This demonstrates the intimate relationship of compact highly dense massive cluster peaks as the main source of the 
Megaparsec tidal force field: filaments should be seen as tidal bridges between cluster peaks. This may be directly understood 
by realizing that a $(-++)$ tidal shear configuration implies a quadrupolar density distribution (eqn.~\ref{eq:quadtide}). In other 
words, the cluster peaks induce a compressional tidal shear along two axes and a dilational shear along the third axis. This 
means that, typically, an evolving filament tends to be accompanied by two massive cluster patches at its tip. These overdense 
protoclusters are the source of the specified shear.

The latter explains the canonical \emph{cluster-filament-cluster} configuration so prominently recognizable in the observed Cosmic Web:  
the nonlinear gravitational evolution results in the typical configuration of filamentary bridges suspended between two or more 
clusters. Figure~\ref{fig:cosmicwebtide} illustrates this close connection between cluster nodes and filamentary bridges. It also illustrates 
the correspondence with the anisotropic force field, as may be inferred from the bars that depict the compressional component 
of the tidal force field. The typical compressional force directed towards the heartline of the filaments is lined up in a  
spatially coherent field of tidal bars along the spine of the filaments, suspended in between the typical tidal configuration around 
cluster peaks. Filaments between two clusters that are mutually aligned show this effect even more strongly. Indeed, these filaments 
turn out to be the strongest and most prominent ones. Studies have also quantitatively confirmed this expectation in $N$-body simulations 
\citep[see][]{colberg05}.

For a full understanding of the dynamical emergence of the Cosmic Web along the lines sketched above, we need to invoke two 
important observations stemming from intrinsic correlations in the primordial stochastic cosmic density field. When restricting 
ourselves to overdense regions in a Gaussian density field we find that mildly overdense regions do mostly correspond to 
filamentary $(-++)$ tidal signatures \citep{pogosyan1998}. This explains the prominence of filamentary structures in the cosmic 
Megaparsec matter distribution, as opposed to a more sheetlike appearance predicted by the Zel'dovich theory. 

The same considerations lead to the finding that the highest density regions are mainly confined to density peaks and their 
immediate surroundings. Following the observation that filaments and cluster peaks are intrinsically linked, the Cosmic Web theory of 
\cite{bondweb96} formulates the theoretical framework that describes the development of the Cosmic Web in the context of
a hierarchically evolving cosmic mass distribution. A prime consideration is that cluster peaks play a dominant dynamical 
role in the Megaparsec force field. This is clearly borne out in the map of the tidal force amplitude in fig.~\ref{fig:cosmicwebtide}. 
This concerns a typical LCDM mass distribution, in this case that of the Millennium simulation \citep{millenium05}. 

\medskip
\cite{bondweb96} argued how clusters, or rather peaks in the primordial density field, can be used to define the bulk of 
the filamentary structure of the corresponding Cosmic Web. They demonstrated how on the basis of a mass ordering of cluster peaks 
in a particular cosmic volume, one may predict in increasing detail the outline of the filamentary spine of the Cosmic Web in that 
same volume. For the accuracy of the predicted web-like structure, it is not sufficient to only take along the location and 
density excess of the primordial peak, but also its shape and orientation. Anisotropic peaks substantially enhance filament formation 
through the quadrupolar gravitational field configuration that they induce. When two such anisotropic peaks are mutually aligned, this effect 
becomes even stronger. In summary, the prominence --- i.e.\ mass surface density and mass --- of the implied filamentary bridges is sensitively dependent 
on a range of aspects of the cluster peaks. The filaments are more prominent and dense when the generating cluster peaks are 
\begin{enumerate}
\item[1.] more massive 
\item[2.] at a closer distance to each other
\item[3.] are strongly anisotropic, i.e.\ elongated, flattened or triaxial
\item[4.] have a favorable mutual orientation, in particular when their long axis is aligned. 
\end{enumerate}

\begin{figure}[!htbp]
\begin{center}
 \includegraphics[width=10.0cm]{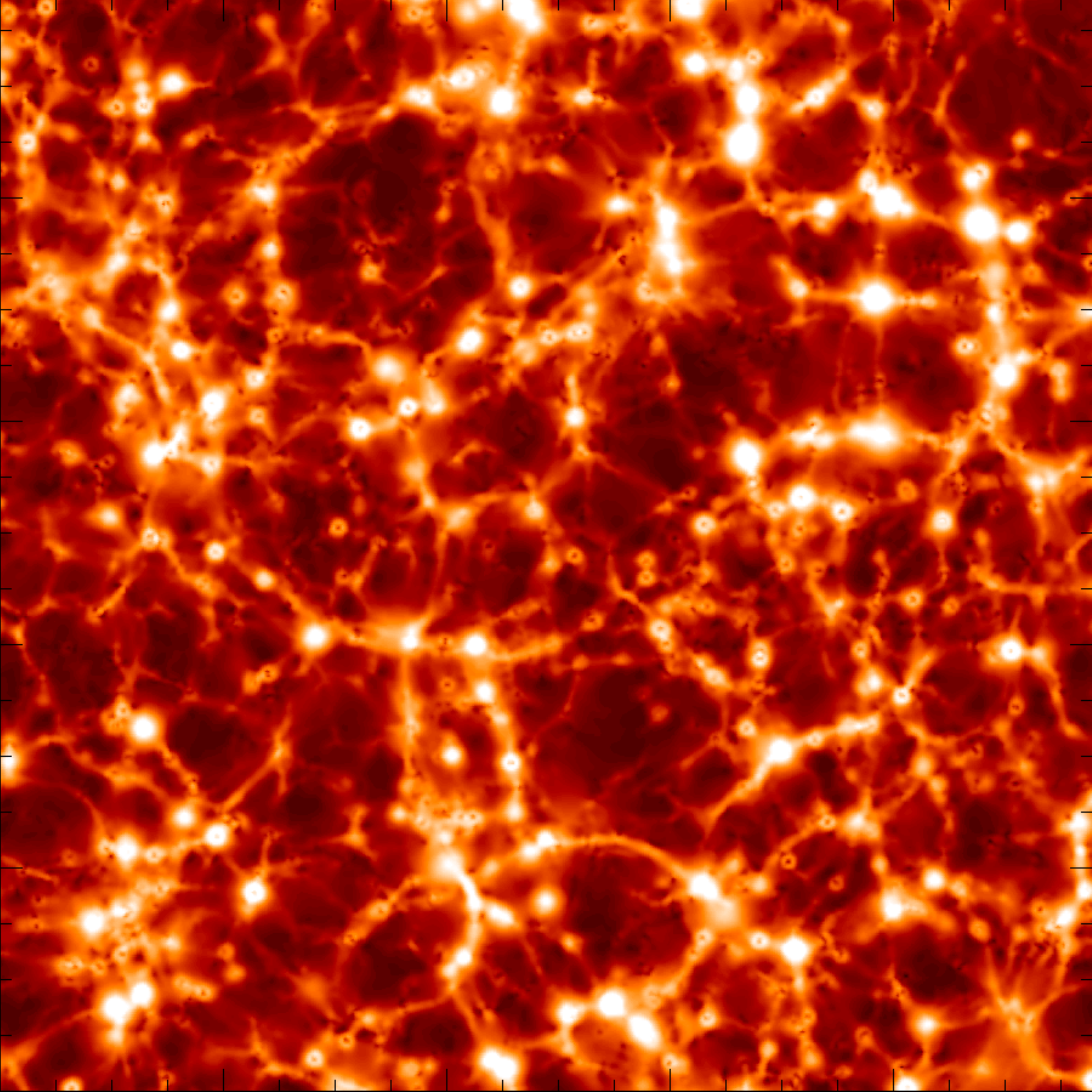} 
 \caption{Clusters and Tidal force field. The image shows the magnitude of the cosmic tidal field, 
filtered on a 1 Mpc/h scale. Clearly visible is the dominance of the clusters in this field, they 
stand out as nodes in the web-like network of relatively tenuous filamentary connections. The 
image concerns the mass distribution in the LCDM cosmology based Millennium simulation. Image 
courtesy: E. Platen.}
\label{fig:mag_tidalfield}
\end{center}
\end{figure}

\section{Cluster-based Reconstruction of the Filamentary Spine}
Based on the above we arrive at the core question of our program. It involves the question whether on the basis 
of a complete or well-selected sample of clusters we may predict, or reconstruct, the filamentary Cosmic Web. 

This question involves a few aspects. One direct one is the concentration on the filamentary spine of the 
Cosmic Web. Another integral anisotropic component of the Cosmic Web are also the walls --- or membranes --- in 
the Cosmic Web. After all, they occupy --- besides void regions --- the major share of the volume in the Universe. 
However, in the observational situation walls are far less outstanding than the filaments. First, filaments 
represent the major share of the mass content in the universe \citep{cautun14}. Secondly, as their matter content is 
distributed over a two-dimensional planar structure, their surface mass density is considerably lower than 
that of the one-dimensional filamentary arteries. In an observational setting this is augmented by an additional 
third effect. \cite{cautun14} showed that the halo mass spectrum is a sensitive function 
of the Cosmic Web environment and has shifted to considerably lower masses in wall-like environments. This 
explains why it is so hard to trace walls in magnitude-limited galaxy surveys: because walls are populated 
but relatively minor galaxies, they tend to be missed in surveys with a magnitude limit that is too high. 
For practical observational applications, we may therefore be forced to concentrate on the filamentary 
aspect of the cosmic mass distribution. 

\begin{figure}[!htbp]
\begin{center}
 \includegraphics[width=13.0cm]{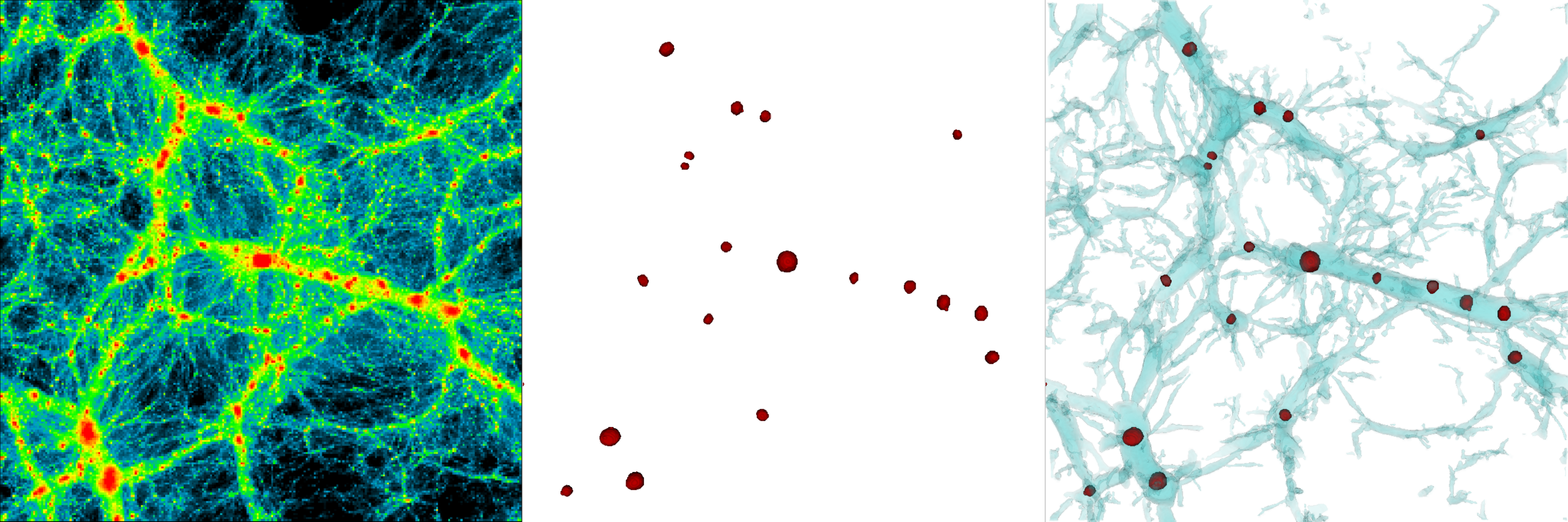} 
 \caption{Aim of filament reconstruction program. Given a given sample of clusters (center), 
constrain the outline of filamentary spine 
of the Cosmic Web (right) and its configuration in the dark matter distribution (left).}
\label{fig:aim}
\end{center}
\end{figure}

Following the considerations outlined above, we arrive at the definition of our program. 
Given cluster locations and properties, we wish to reconstruct their surroundings and study the Cosmic Web 
as defined by the clusters. The program entails the following principal target list (see fig.\ref{fig:aim}). 
\begin{enumerate}
  \item[1.] Use Cosmic Web theory to predict the location of filaments given cluster observations including shape, 
size and mutual alignment.
  \item[2.] Predict the prominence and other intrinsic characteristics of filaments. 
  \item[3.] Investigate the application to the observational reality, taking into account systematic biases and effects, and 
observational errors. 
\end{enumerate}

\subsection{Observational context and potential}
The observational significance of our program is based on the realization that, beyond a rather limited depth, it will 
be far more challenging to map the filamentary features in a cosmic volume than it is to map its cluster population. 
While filaments are well-known features in galaxy surveys of the nearby Universe, such as 2dFGRS and SDSS, it is hard 
to trace them at larger depths. Moreover, mapping the outline of the dark matter Cosmic Web will be almost beyond 
any practical consideration, although \citep{dietrich12} recently managed to trace a strong intercluster filament. On the other hand, 
the cluster distribution is or will be accessible throughout large parts of the observable universe. Armed with the ability to 
reconstruct the Cosmic Web given the observed cluster constraints, we would be enabled to explore the possibility to 
study its characteristics or observe filaments and investigate their properties in targeted campaigns. To this end, it is possible 
to probe the cluster distribution in different wavelength regimes. Most promising are the cluster samples that will be obtained from upcoming 
X-ray regime surveys, of which in particular 50k-100k cluster sample of the eROSITA all-sky survey will be an outstanding example 
\citep{merloni12}. Given the implicit relation to the tidal force field, major gravitational weak lensing surveys --- such at present the 
KIDS survey an the future Euclid survey --- will certainly open the possibility to explore the dark matter Cosmic Web in considerably 
more detail, and relate its characteristics to the observed galaxy distribution. Perhaps even most promising will be upcoming 
observations of the Sunyaev-Zel'dovich effect, which will potentially allow a volume-complete sample of clusters throughout the 
observable universe. 

\firstsection
\section{BARCODE Reconstruction Formalism}
There are several methods that allow the reconstruction of a full density field based on incomplete sampled information. 
The first systematic technique for allowing the inference on the basis of a restricted set of observational constraints 
is that of the constrained random field formalism \citep{edb87,hofrib91,weyedb96}, and its implementation 
in observational settings involving measurement errors in which the Wiener filter is used to compute the mean field 
implied by the observational data \citep{bond95,zaroubi95}. However, strictly speaking it is only suited for Gaussian 
random fields, so that in a cosmological context it is merely applicable to configurations that are still --- nearly --- 
in the linear regime of structure formation \citep{BBKS86,adler1981,adler2007}. 

For the project at hand we seek to translate constraints based on a set of cluster observations into a density field. 
Clusters, however, are highly nonlinear objects. It is extremely challenging to relate an evolved cluster to the 
primordial density peak out of which it evolved. This is a consequence of the complex hierarchical buildup of a cluster. 
A cluster emerges as the result of the assembly of a few major subclumps, augmented by the infall of numerous small clumps 
and the influx of a continuous stream of mass, along with the gravitational and tidal influence of surrounding matter 
concentrations \citep[see][for an insightful study]{porcludl2011}. Any attempt to relate a present-day cluster directly 
to a unique primordial density peak is doomed. 

We therefore follow a ``from the ground up'' approach, in the context of a Bayesian reconstruction formalism. Besides 
allowing us to find statistically representative primordial configurations that would relate to the observed cluster 
configurations that emerged as a result of fully nonlinear evolution, it has the advantage of enabling the 
self-consistent incorporation of any number of additional physical and data models. In recent years, several groups 
have worked on the development and elaboration of sophisticated and elaborate Bayesian inference techniques. Original 
work along these lines by \citet{kitaura08} culminated in the development of KIGEN \citep{kitaura12}, on which the 
present \barcode\ formalism is based. KIGEN allows the translation of the observed distribution of galaxies, such as 
e.g.\ those in the 2MRS survey, into a realization of the implied primordial density and velocity field in the 
sample volume \citep{kitaura12,hesskitaura13}. Subsequent gravitational evolution reveals the implied nonlinear structure in the 
dark matter distribution. Along the same lines, elaborate Bayesian reconstruction formalisms have been developed 
by Jasche, Wandelt, Leclercq and collaborators \citep{jaschewandelt13,leclercq2015b}.  The latter allowed even the 
study of the void population in the dark matter distribution implied by the SDSS survey \citep{leclercq15}.  

While the Bayesian formalism is basically oriented towards obtaining constrained samples of the 
primordial Gaussian density and velocity field underlying the measured observational reality that 
was imposed as constraint, the intention of our \barcode\ program concerns the Cosmic Web that 
is emerging as a result of the gravitational evolution of structure. To this end, the primordial field 
sampled from the posterior probability distribution form the initial conditions for the subsequent 
gravitational evolution. The latter is followed by means of an $N$-body code or of an analytical structure 
formation model. In our case, we may analyze the structure of the various structural components of the 
emerging Cosmic Web. 

\subsection{Bayesian formalism}
The basic philosophy of Bayesian reconstructions is to find a set of realizations that form a statistically 
representative sample of the posterior distribution of allowed models or configurations given the (observed) 
data sample. In essence, Bayesian inference is about finding the model that best fits the data, while taking into 
account all prior knowledge. Philosophically, it is based on the interpretation of the concept of probability in 
terms of the degree of belief in a certain model or hypothesis \citep{neal93,jaynes2003probability}. 

\medskip
Central to Bayesian inference is Bayes' theorem, which formalizes the way to update our a priori belief $\prior(m) = P(m)$ in 
a model $m$ into an a posteriori belief $P(m|d)$, by accounting for new evidence from data $d$, 
\begin{equation}
\label{eqn:bayes_theorem}
  P(m|d) = \frac{P(d|m) P(m)}{P(d)} \equiv \frac{\likeli(d|m) \prior(m)}{P(d)} \,, 
\end{equation}
in which we recognize the following terms and probabilities, 

\medskip
\begin{enumerate}
  \item the \emph{prior} $\prior(m) = P(m)$: the probability distribution that describes our a priori belief in the models $m$, without regard for the observed data;
  \item the \emph{likelihood} $\likeli(d|m) = P(d|m)$: the probability function that describes the odds of observing the data $d$ given our models;
  \item the \emph{posterior} $P(m|d)$: the updated belief in our models, given the data;
  \item and the \emph{evidence} $P(d)$: the chance of observing the data without regard for the model.
        In Bayesian inference this term can be ignored, as we are only interested in the relative probabilities of different models, meaning that the evidence always factors out.
\end{enumerate}

\medskip
In the context of our Bayesian reconstruction formalism, the data $d$ are the constraints inferred or processed from the 
raw cluster observations. As we will argue below, in sect.~\ref{sec:constraints}, there are several means in which galaxy or cluster 
observations can be imposed as constraints. In order to facilitate the imposition of heterogeneous cluster data, 
the \barcode\ algorithm has chosen for a representation in terms of values on a regular grid in Eulerian space. 

The data are specified at Eulerian positions $\{ \bx \}$. Lagrangian coordinates $\{ \bq \}$ and Eulerian coordinates 
are related via the displacement field $\displace({\bq})$, 
\begin{equation*}
    \bx(\bq) = \bq + \displace(\bq) \,.  
\end{equation*}
The displacement field entails the dynamics of the evolving system. 

\subsection{the \barcode\ constraints}
\label{sec:constraints}
The formalism that we develop with the specific intention of imposing a set of sparse and arbitrarily located cluster constraints is called \barcode.
It is an acronym for {\sc Ba}yesian {\sc R}econstruction of {\sc CO}smic {\sc DE}nsity fields. 

In \barcode\ we have translated the observational (cluster) data at a heterogeneous sample of Eulerian locations 
$\{ \bx_G \}$ into a regular image grid $\{\rho^\obs(\bx)\}$, represented on a regular Eulerian grid. A crucial ingredient to 
facilitate this transformation is the use of a mask. Given the sparseness and heterogeneous nature of 
the imposed cluster constraints, the formalism includes a mask to indicate the regions that have not, incompletely or 
selectively covered. 

It is in particular on the aspect of using an regular grid $\{\rho^\obs(\bx)\}$ to transmit the observational 
constraints that \barcode\ follows a fundamentally different approach from the one followed in the related 
KIGEN formalism \citep{kitaura12}. KIGEN is able to process the galaxy locations in the dataset 
directly for imposing the data constraints. It involves two stochastic steps instead of one. It first 
samples a density field given the galaxy distribution. In a second step it then samples a new particle distribution, 
given the sampled density field and a model of structure formation. It renders KIGEN highly flexible. 
It does require the posterior probability function to be differentiable, and is able to implicitly 
take into account the bias description for the galaxy sample. 

A grid-based formalism like \barcode, on the other hand, requires an explicit bias description. Also, it requires 
the posterior to be differentiable, because of the use of the Hamiltonian Monte Carlo HMC sampling procedure 
(see sect.~\ref{sec:hmc}). 
 
\subsection{the signal: \barcode\ samples}
In the context of the Bayesian formalism, the signals that \barcode\ is sampling are the models constrained 
by the data. In our application they are realizations of the primordial density field $\{\delta(\bq)\}$. The 
core task of the reconstruction formalism is therefore the proper sampling of a realization of the 
primordial density field $\{\delta(\bq)\}$, given the posterior distribution function for $\{\delta(\bq)\}$. 
The posterior distribution function is determined on the basis of the input data and the assumed cosmological model. 

The key product of the procedure is therefore a sample $\delta(\bq)$ of the \emph{primordial density fluctuation field}. 
For practical computational reasons, the primordial field is specified at the Lagrangian locations $\{ \bq \}$ on a 
regular grid. In our test runs, we used a grid with $N_x = 64$ cells on a side for this discrete representation of an 
intrinsically fully continuous field on a finite regular grid. The presently running implementations have $N_x=128$ 
or $N_x=256$ cells. Each of the density values $\delta(\bq_1)$ is one of the stochastic variables to be sampled. The 
total signal ${\cal S}$,  
\begin{equation}
  {\cal S} = \left\{ \delta(\bq_1), \delta(\bq_2), \dots, \delta(\bq_N) \right\} = \{\delta(\bq_i)\} \,,
\end{equation}
defines a $D=N_x^3$ dimensional probabilistic space. In other words, for $N_x=64$ we are dealing with 
the sampling of a full Bayesian signal in a 262144-dimensional probabilistic space.

\subsection{the \barcode\ Model Prior}
\label{sec:prior}
The only thing we know about the primordial density field --- without regard for the data --- is that it must be a 
Gaussian random field; this should fully characterize our prior belief in such a field. The prior in our model is 
therefore a multivariate Gaussian distribution characterized by a 
zero mean and a correlation matrix determined by the correlation function as measured from the cosmic microwave background:
\begin{equation}
  \prior(\{\delta(\bq)\} | P(k)) = \frac{1}{\sqrt{{(2\pi)}^N \lVert S \rVert}} \exp\left( - \frac12 \sum_{i,j} \delta(\bq_i) 
{\xi^{-1}}_{ij} \delta(\bq_j) \right)\,,
\end{equation}
where $\xi^{-1}_{ij}$ is the inverse of the correlation matrix $\xi_{kl}$, the real-space dual of the Fourier-space power spectrum $P(k)$ with elements
\begin{equation}
\xi_{kl} = \xi(\lvert \bq_k - \bq_l \rvert),. 
\end{equation}

The cosmological model underlying our reconstruction enters via the cosmological power spectrum $P(k)$. It contains implicit 
information on global cosmological parameters, on the nature of the primordial density fluctuations, and via these the nature 
of dark matter. 

\medskip
In addition to the primordial fluctuations, we also need to specify the structure formation model, which transforms 
the Lagrangian to Eulerian space density fields via the transformation $f:\delta(q) \to \rho(x)$ from a Lagrangian space 
overdensity field $\delta(q)$ to Eulerian space density field $\rho(x)$. In the context of gravitationally driven 
structure formation, one could employ an $N$-body simulation 
to optimally follow the structure formation process, including the complex nonlinear aspects. However, in practice it 
would be unfeasible to include this in the formalism, as it would take an overpowering computational effort to 
numerically compute derivatives with respect to the signal ${\cal S}$.

For the computation of the displacement, \barcode\ therefore prefers to use analytical prescriptions. The most 
straightforward option is the Zel'dovich formalism. Other options are higher-order Lagrangian perturbation 
schemes such as 2LPT or ALPT (ALPT, or Augmented Lagrangian Perturbation Theory, is a structure formation 
model that combines 2LPT on large scales with spherical collapse on small scales).

Closely related to the latter is a model $f:\bx\rightarrow\zspace$ that subsequently transforms Eulerian coordinates 
to redshift space coordinates, and hence transforms the modeled density field from Eulerian space to redshift space for 
the situations in which the observational data are in fact redshift data. This is implicitly included in the structure 
formation model via its prescription for the corresponding velocity perturbations.           

In addition to these cosmological and structure formation models, we also need to take along model descriptions 
relating the observed galaxy or cluster distribution to the underlying distribution. This includes factors such as 
a noise model for statistical sources of error, like measurement uncertainties in the observations, and a bias 
model connecting the galaxy/cluster distribution to the density field. 

\subsection{MCMC sampling and the Hamiltonian Monte Carlo Procedure}
\label{sec:mcmc}
On the basis of the set of data constraints, and the model assumptions discussed in the previous section, we obtain 
the posterior distribution function 
\begin{equation}
P({\cal S})=P(\left\{ \delta(\bq_1), \delta(\bq_2), \dots, \delta(\bq_N) \right\})\,.
\end{equation}
To obtain meaningful constrained realizations for our problem, we sample the posterior distribution. Given the complex 
nature of the posterior distribution function in an $D=N_x^3$ dimensional probability space, an efficient sampling 
procedure is crucial in order to guarantee a statistically proper set of samples ${\cal S}$. 

Markov chain Monte Carlo (MCMC) methods try to alleviate the computational costs of sampling a high-dimensional probability 
distribution. They do this by constructing a so called Markov chain of samples $s_i$. In our case, this becomes a chain 
of density fields, each sampled on a regular grid. Such a chain may be seen as a random walk through the parameter space 
of the pdf $P(\{\delta(\bq)\})$. Each visited location in the high dimensional space corresponds to one particular 
sample realization of the density field $\delta(\bq)$. 
To do so, MCMC chains automatically visit high probability regions in direct proportion to their relative probability, and do 
not waste computational tie on low probability areas. The performance gain can be orders of magnitude for high-dimensional problems.
The resulting sampling ensemble is equivalent to an ensemble from the full posterior $P(\{\delta(\bq)\})$. 

Arguably the best known MCMC method is the Metropolis-Hastings sampler. Another well-known algorithm is the closely 
related concept of Gibbs sampling, basically a multi-dimensional extension of Metropolis-Hastings algorithm.  
In \barcode\ we follow a far more efficient methodology, that of the Hamiltonian Monte Carlo sampling. 

\subsection{Hamiltonian Monte Carlo sampling --- HMC}
\label{sec:hmc}
Hamiltonian sampling is based on a physical analogy, and uses concepts from classical mechanics --- specifically Hamiltonian dynamics --- to solve the statistical sampling problem \citep{neal93,neal12,jaschekitaura10} (also see \cite{leclercq2015b}
for an excellent summary of the ideas involved). Following the concept of Hamiltonian mechanics, a Markov chain is produced 
in the form of an \emph{orbit} through parameter space, which is found by solving the Hamiltonian equations of motion:
\begin{eqnarray}
\label{eqn:eom_general}
  \dif{\posHMC_i}{\tau} &= {\displaystyle \partial {\hamHMC} \over \displaystyle \partial {\momHMC_i}} \nonumber\\
  \dif{\momHMC_i}{\tau} &= -{\displaystyle \partial {\hamHMC} \over \displaystyle \partial {\posHMC_i}} \,,
\end{eqnarray}
where $\tau$ is an artificial ``time'' analogue that we merely use to evolve our Markov chain. Important 
in this respect is also the time reversibility of Hamiltonian dynamics. The positions $\posHMC_i$, the 
momenta $\momHMC_i$ and the Hamiltonian $\hamHMC$, the sum of a potential term $\potHMC$ and a kinetic term $\kinHMC$, 
\begin{equation}
\hamHMC = \potHMC + \kinHMC \,,
\label{eqn:hamiltonian}
\end{equation}
are not real physical quantities, but equivalent stochastic variables. It is straightforward to interpret the meaning of the position vectors $\posHMC_i$.
They represent the primordial density field realization $\delta(\bq)$, and the dynamical orbit the chain of density field samplings,
\begin{equation}
\posHMC_i=\delta(\bq_i)\,.
\end{equation}
In order to avoid confusion, notice that the stochastic coordinate analogues $\posHMC_i$ are something completely 
and fundamentally different than the Lagrangian coordinates $\bq$. The potential term $\potHMC$ is identified with 
the probability distribution function $P$, and is equal to its negative logarithm, $\potHMC= -\log P$. 
The $D$-dimensional auxiliary momentum vector $\momHMC_i$ is related to the steepness of the change in density field 
sample between sampling steps. The kinetic energy term $\kinHMC$ is the quadratic sum of these auxiliary momenta,  
\begin{equation}
  \kinHMC = \frac12 \momHMC^T \masHMC^{-1} \momHMC \,.
  \label{eqn:kinetic_term}
\end{equation}
in which the symmetric positive definite mass matrix term $\masHMC$, called mass in analogy to the situation in 
the context of classical mechanics, characterizes the inertia of the sampling orbit moving through parameter space. 
It is of crucial importance for the performance of the formalism: if the mass term is too large it will result in 
slow exploration efficiency, while masses which are too light will result in large rejection rates. 

\begin{figure}[t]
\begin{center}
\mbox{\hskip -1.0cm \includegraphics[width=13.0cm]{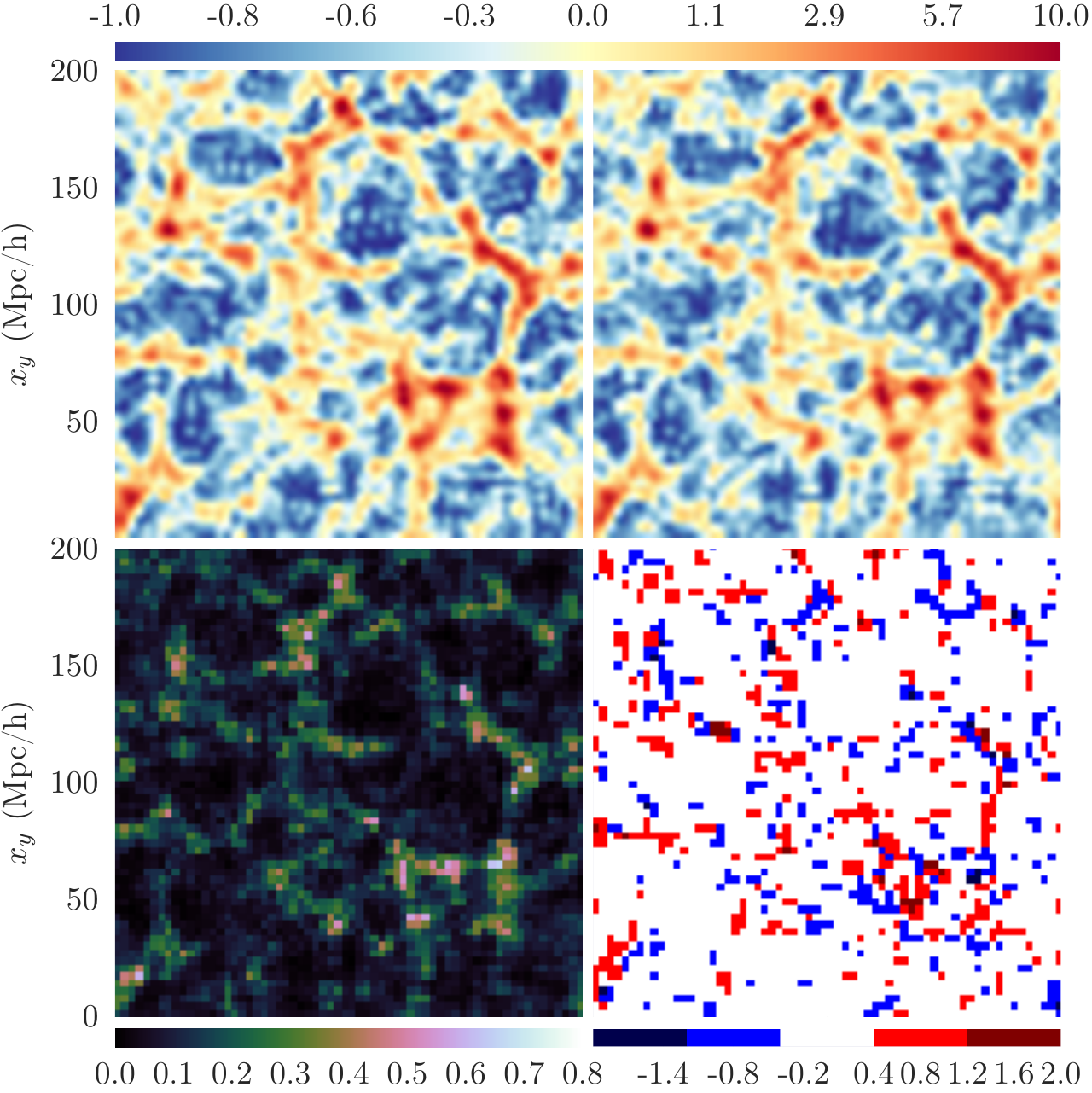}} 
 \caption{Comparison of true to sampled \emph{Eulerian} density fields.
  \emph{Top left:} true.
  \emph{Top right:} mean of 60 samples (iteration 500, 550, \dots, 3500) after burn-in phase.
  \emph{Bottom left:} standard deviation of 60 samples.
  \emph{Bottom right:} difference of true and mean fields.
>From E.G.P. Bos 2016}
\label{fig:density_fields}
\end{center}
\end{figure}

\medskip
As in the case of other MCMC algorithms, for the startup of the procedure there is an additional important 
aspect, the \emph{burn-in} phase. An MCMC chain starts up from an initial guess for the signal ${\cal S}$ to 
be sampled, and subsequently evolves a chain from there. As we may not necessarily start from a region of 
high probability, it may take the MCMC chain a series of iterations before reaching this region. The 
period up to this point is called the  burn-in phase of the chain. The samples obtained in the burn-in phase 
are discarded from the ensemble, since they are not part of a representative sample.

\section{BARCODE validation: Cosmic Web reconstruction}
In \barcode\ we have implemented an HMC sampler of primordial density fields $\delta(q)$ given density data 
from observations. The algorithm can deal both with real-space density fields, as well as density fields 
sampled in redshift space. To incorporate a structure formation model, we can use either the Zel'dovich approximation 
or an ALPT structure formation model, which combines the higher-order 2LPT Lagrangian model on large scales with 
spherical collapse on small scales~\citep{hesskitaura13}. The implementation may invoke a Gaussian or 
Poisson error model for the observed densities. 

\begin{figure}[t]
\begin{center}
\mbox{\hskip -0.0cm \includegraphics[width=13.5cm]{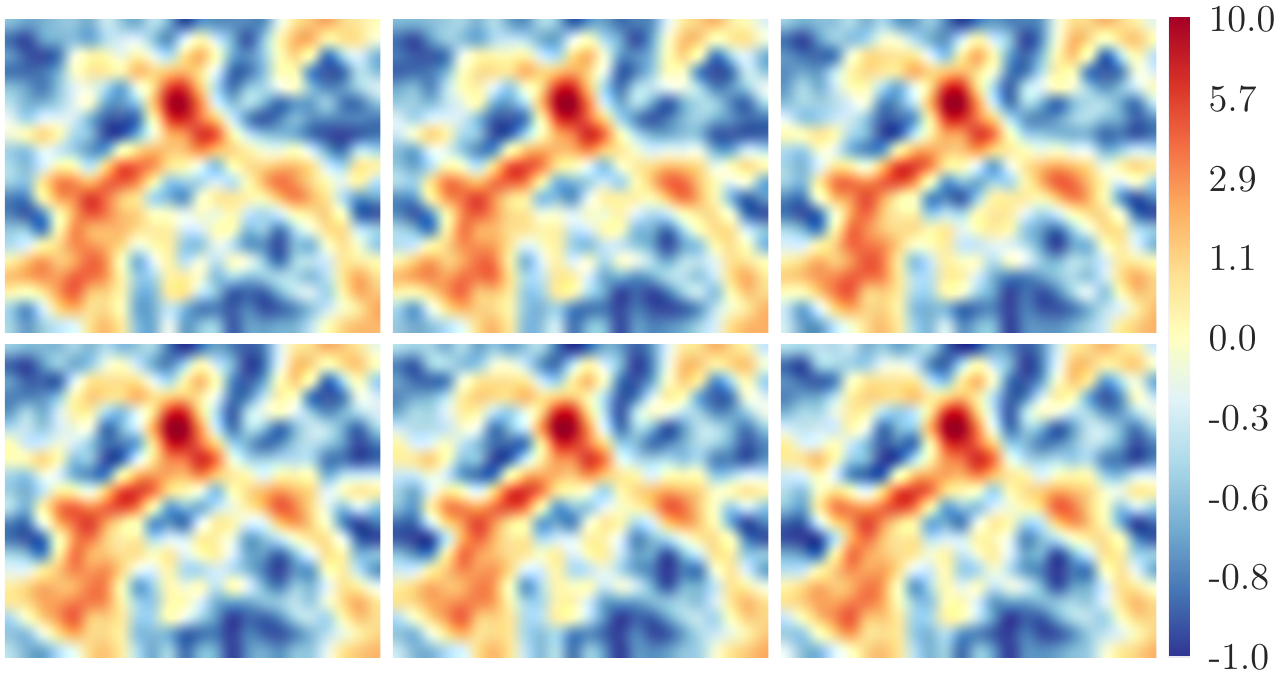}} 
 \caption{Zoom-ins of  Eulerian space density field samples at iterations 500, 1000, 1500, 2000, 2500 and 3000 (left to right). 
>From E.G.P. Bos 2016}
\label{fig:density_fields_zoom}
\end{center}
\end{figure}

\begin{figure}[b]
\centering
\includegraphics[width=13.5cm]{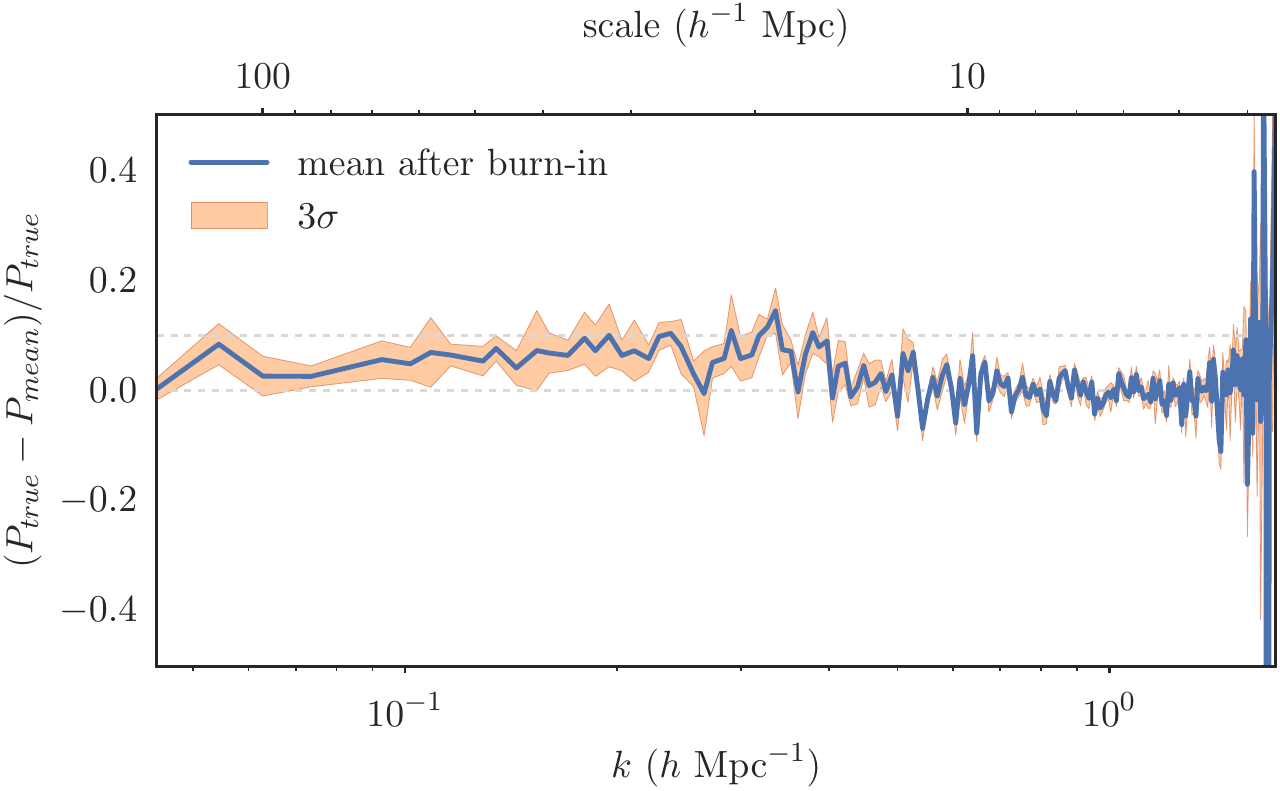} 
  \caption{Relative differences between true and mean Lagrangian power spectra for 60 post-burn-in samples. 
    The shaded area represents (three times) the standard deviation of the power spectra of the  samples.
>From E.G.P. Bos 2016}
\label{fig:powerpk}
\end{figure}

\subsection{Test results}
Here we first present the result of some numerical experiments, demonstrating the potential to reconstruct successfully 
the web-like outline of the cosmic matter distribution, given an observed sample of the large scale 
density distribution. 

As a test, we have evaluated the performance of our code on a LCDM dark matter simulation, in which the 
matter distribution in a box with a side of $L = 200 \mpch$ is represented on a grid of $N_x = 64$ cells. 
It means the stochastic sampling process takes place in a $262144$-dimensional space. We find that the 
algorithm starts constraining the large scales (low modes) and then increasingly gains power towards 
small scales (high modes). 

Following the burn-in phase, the chain remains in the high probability region around the ``true'' density.
The sampling chain is expected to vary around this true density due to the random kicks from 
the (statistical, auxiliary) momentum. Figure~\ref{fig:density_fields} shows the comparison between 
the Eulerian space mean density field of 60 samples --- obtained from 3000 iterations --- and the true 
underlying density field. The match between the two is very good. We notice this on account of a visual 
comparison between the matter distribution in the top frames. This is strongly supported by the quantitative 
comparison in terms of the standard deviation of the 60 samples (bottom left) and the 
difference field between the true and mean fields (bottom right-hand plot). 

The differences between mean and sampled field turn out to be rather subtle. In Eulerian space they tend 
to be most prominent in the high-density regions. In part, this can be understood from the realization that 
there is more variation in those regions in combination with the fact that there are not enough samples 
to completely average out these variations. 

To appreciate the variation between field variations, in figure~\ref{fig:density_fields_zoom} we zoom in 
on a couple of post-burn-in iterations. It shows that differences are hard to discern, in particularly at 
this low resolution. At a careful inspection we may notice some minor differences.
For instance, the height of the two large peaks vary significantly. Also, the shape of the ``cloud'' in the void, 
just on the lower right from the center, varies from almost spherical with a hole in the middle to 
an elongated and even somewhat curved configuration. 

Yet, overall the reconstructions appear to be remarkably good. This is true over the entire range of scales 
represented in the density field. The power spectra of the chain samples properly converge to the true power 
spectrum, as may be inferred from figure~\ref{fig:powerpk}.

\section{Reconstructions in Redshift Space}
One important caveat in the basic \barcode\ algorithm is that in the observational practice we 
tend to use redshifts as measures for distance. Because redshifts also include the contribution 
from peculiar motions, the measured distribution does not concern comoving space $\bx$, but the 
distribution in redshift space $\zspace$. 

The direct implication is that our data will be deformed by so-called redshift space distortions.
Clusters will seem radically stretched along the line of sight due to the internal virialized 
velocities of their galaxies. On the other hand, large scale overdense structures appear squashed 
along the line of sight, while underdense structures are observed to be larger in the radial direction 
than in reality. It is one of the principal reasons for the appearance of \emph{Great Walls} 
perpendicular to the line of sight. All these effects must be taken into account in order to accurately 
measure volumes, densities, shapes and orientations of the components of the Cosmic Web.

There are several options to accommodate redshift space distortions when processing 
measurements. Most of these involve a form of a posteriori or a priori ``correction'' of 
the redshift space effects. A priori corrections include, for instance, the detection of 
``Fingers of God'', followed by an automatic reshaping into spheres. Possible a posteriori 
corrections include the reshaping of the power spectrum to account for the ``Kaiser effect'' 
of squashing along the line of sight. However, all these methods are approximations and most 
of them focus on one aspect of the redshift space transformation only.

We opt for a self-consistent redshift space correction in the \barcode\ formalism. This 
concerns a one-step transformation of \emph{data model} itself into redshift space. To 
the best of our knowledge, this is the first self-consistent redshift space data based 
reconstruction of the primordial density fluctuations. To this end, we adjust our \barcode\ 
data model to not only transform from Lagrangian coordinates $\bq$ to Eulerian coordinates 
$\bx$ using a structure formation model, but subsequently also from $\bx$ to the corresponding 
redshift space. 

Since our structure formation formalisms are particle based (using Lagrangian perturbation theory), 
this can easily be achieved by means of one more coordinate mapping from Eulerian space to redshift space.
In this way, we incorporate automatically and directly all possible redshift space effects in our 
reconstructions, insofar as the structure formation model includes them. The one additional 
complication for such a strategy in the context of the Hamiltonian Monte Carlo formalism, 
is that for the redshift space transformation to work it must facilitate analytical derivatives 
of the posterior distribution with respect to the sampled fields. To this end, we have 
derived and implemented the required equations in the \barcode\ code. 

\subsection{Mapping from Eulerian to Redshift space}
Redshift space coordinates $\zspace$ can be defined as
\begin{equation}
  \zspace \equiv \bx + \frac{1}{Ha} \left( (\bv-\bv_\obs) \cdot \hatx \right) \hatx \equiv \bx + \displacez \,,
\label{eqn:redshift_space}
\end{equation}
where $\zspace$ is a particle's location in redshift space, $\bx$ is the Eulerian comoving coordinate, $\bv$ is the 
particle's (peculiar) velocity and $\bv_\obs$ is the velocity of the observer.
$a$ is the cosmological expansion factor, which converts from physical to comoving coordinates, and $H$ is the 
Hubble parameter at that expansion factor, which converts the velocity terms to distances with units $\mpch$. 
We call $\displacez$ the \emph{redshift space displacement field}, analogously to the displacement field 
$\displace$ that transforms coordinates from Lagrangian to Eulerian space in Lagrangian perturbation theory. 

An important point to keep in mind is that in equation~\ref{eqn:redshift_space}, the comoving coordinates are defined 
such that the observer is at the origin. The unit vector $\hatx$, and thus the entire redshift space, changes completely 
when the origin shifts. 

\begin{figure}[b]
\centering
\mbox{\hskip -0.5cm \includegraphics[width=14.5cm]{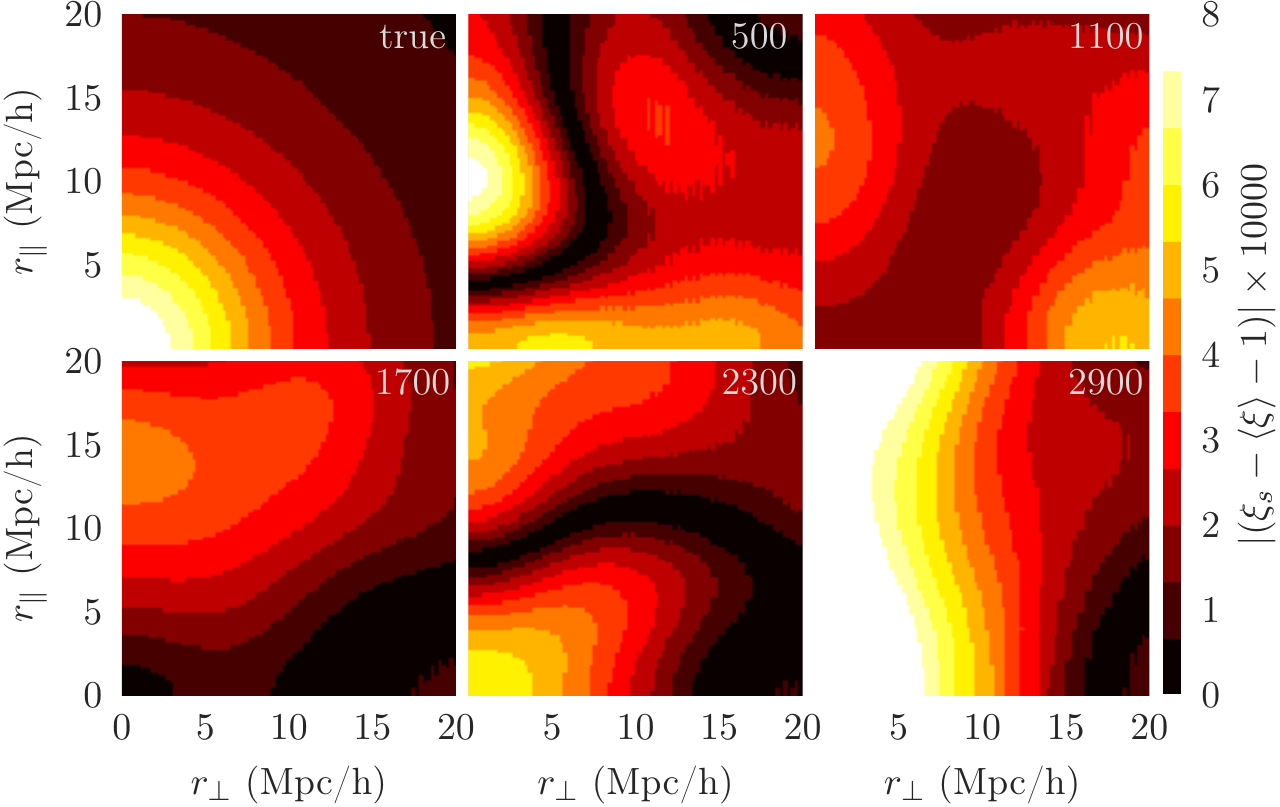}} 
 \caption{Five sample Eulerian 2D correlation functions minus the mean
of all reconstructions, compared to the true field in the top left panel.}
\label{fig:zspace}
\end{figure}

\subsection{Test results: redshift space correlation functions}
Overall, in our $64^3$ test-runs based on redshift space renderings of the simulations described 
in the previous section we found that an excellent performance. It manages to deal with all 
redshift space artifacts, and differences between the sampled density field and the true 
density field remains limited to small local variations. This is confirmed by the power 
spectrum, which seems to match the expected true power spectrum in all necessary detail. 
Our \barcode\ formalism has fully and self-consistently accounted for the Kaiser 
redshift space effect and even additional mildly nonlinear effects such as in the 
triple value region corresponding to the infall region around clusters. 

Perhaps the most convincing demonstration of this success is that of the 2D 
correlation function $\xi(\pi,\sigma)$. It measures the correlation function 
by segregating the contributions along two direction one along the line of sight, 
the other perpendicular. In the case of an intrinsic anisotropies in the data, we 
will obtain a deviation from a perfectly circular $\xi(\pi,\sigma)$. 

As for the \barcode\ reconstructions, the two-dimensional correlation functions 
of the reconstructions match the true --- almost isotropic and perfectly circular --- 
one when using the redshift space model. 
In other words, it shows that it manages to eliminate large scale redshift
space anisotropies. Figure~\ref{fig:zspace} shows the differences of the 
Eulerian 2D correlation functions $\xi(\pi,\sigma)$ of five 
\barcode\ samples  compared to that for the true field. The differences 
are very small, in the order of $\delta \xi/\xi \approx 10^{-3}$. Nonetheless, 
we may note some variation in the differences. This is a result of a minor 
effect, a result of the MCMC sampling process itself introducing additional 
anisotropies in the samples

In summary, we may conclude that the \barcode\ formalism has successfully 
incorporated the ability to operate directly in redshift space.


\begin{thebibliography}{999}

\bibitem[Adler (1981)]{adler1981}
	Adler R.J. 1981, 
        \textit{The Geometry of Random Fields} (Wiley)

\bibitem[Adler \& Taylor (2007)]{adler2007}
	Adler R.J. \& Taylor J.E. 2007, 
        \textit{Random Fields and Geometry} (Springer)

\bibitem[Bardeen et al. (1986)]{BBKS86}
	Bardeen J.M., Bond J.R., Kaiser N. \& Szalay A.S. (BBKS) 1986,
	\textit{ApJ},	~304, 	15
	
\bibitem[Bertschinger (1987)]{edb87}
	Bertschinger E. 1987, 
	\textit{ApJL}, ~323, 103

\bibitem[Bond (1995)]{bond95}
	Bond J.R. 1995,
	\textit{Phys Rev. Lett.}, 74, 4369

\bibitem[Bond et al. (1996)]{bondweb96}
	Bond J.R., Kofman L. \& Pogosyan D. 1996,
	\textit{Nature}, ~6575, 603

\bibitem[Bond et al. (1998)]{bondweb1998}
        {Bond} J.R., {Kofman} L., {Pogosyan} D. \& {Wadsley} J. 1998,
        in S. Colombi, Y. Mellier eds., 
        \textit{Wide Field Surveys in Cosmology, 14th IAP meeting) (Editions Frontieres}, p. 17

\bibitem[Bos (2016)]{bos2016}
	Bos E.G.P.. 2016,
        \textit{Voids, Clusters and Reconstructions of the Cosmic Web}, PhD thesis, Univ. Groningen

\bibitem[Cautun et al. (2014)]{cautun14}
	Cautun M., van de Weygaert R., Jones B.J.T., \& Frenk C.S.  2014, 
	\textit{MNRAS},	~441, 	2923

\bibitem[Colberg et al. (2005)]{colberg05}
         {Colberg} J.M., {Krughoff} K.S. \& {Connolly} A.J. 2005, 
         \textit{MNRAS}, 359, 272

\bibitem[Colless et~al. (2003)]{Col03} 
        {Colless} M., et al. 2003, 
        \textit{ArXiv e-prints}, 0306581 

\bibitem[Dietrich et al. (2012)]{dietrich12}
        {Dietrich}, J.~P., {Werner}, N., {Clowe}, D., {Finoguenov}, A., 
        {Kitching}, T., {Miller}, L. \& {Simionescu}, A. 2012,
        \textit{Nature}, 487, 202

\bibitem[Guzzo et al. (2013)]{guzzo2013}
	Guzzo L. \& The VIPERS team 2013, 
	\textit{The Messsenger}, ~151, 41

\bibitem[Guzzo et al. (2014)]{guzzo2014}
	Guzzo L., et al. 2014, 
	\textit{AA}, ~566, 108

\bibitem[Hess et al. (2013)]{hesskitaura13}
	Hess S., Kitaura F.-S. \& Gottl\"ober S. 2013, 
	\textit{MNRAS}, ~435, 2065

\bibitem[Hoffman \& Ribak (1991)]{hofrib91}
	Hoffman Y. \& Ribak E. 1991, 
	\textit{ApJL}, ~380, 5

\bibitem[Huchra et al. (2012)]{HMM12}
	Huchra J.P., et al. 2012, 
	\textit{ApJS}, ~199, 26

\bibitem[Jasche \& Kitaura (2010)]{jaschekitaura10}
	Jasche J. \& Kitaura F.-S. 2010,
	\textit{MNRAS}, ~407, 29

\bibitem[Jasche \& Wandelt (2013)]{jaschewandelt13}
	Jasche J. \& Wandelt B.-D. 2013, 
	\textit{MNRAS}, ~432, 894

\bibitem[Jaynes (2003)]{jaynes2003probability}
	Jaynes E.T. 2003,
	\textit{Probability Theory: The Logic of Science} (Cambridge Univ. Press)
 
\bibitem[Kitaura \& Ensslin (2008)]{kitaura08}
	Kitaura F.-S. \& Ensslin T.A. 2008,
	\textit{MNRAS}, ~389, 497

\bibitem[Kitaura (2012)]{kitaura12}
	Kitaura F.-S. 2012,
	\textit{MNRAS}, ~429L, 84

\bibitem[Leclercq et al. (2015)]{leclercq15}
	Leclercq F., Jasche J., Sutter P.M., Hamaus N. \& Wandelt B.  2015,
        \textit{JCAP}, ~03, 047

\bibitem[Leclercq (2015)]{leclercq2015b}
	Leclercq F. 2015,
        \textit{Bayesian large-scale structure inference and cosmic web analysis}, 
        PhD thesis, Univ. Pierre et Marie Curie, Institut d'Astrophysique de Paris

\bibitem[Ludlow \& Porciani (2011)]{porcludl2011}
	Ludlow A.D. \& Porciani C. 2011,
        \textit{MNRAS}, 413, 1961

\bibitem[Merloni et al. (2012)]{merloni12}
        Merloni, A. et al. \& the German eROSITA Consortium 2012,
        \textit{ArXiv e-prints}, 1209.3114

\bibitem[Neal (1993)]{neal93} 
         Neal R.M. 1993,
         \textit{Probabilistic Inference Using Markov Chain Monte Carlo Methods},
         Technical Report CRG-TR-93-1, Dept. Comp. Science, Univ. Toronto

\bibitem[Neal (2011)]{neal12}
	Neal R.M. 2011, 
	\textit{MCMC using Hamiltonian dynamics}, 
        in S. Brooks, A. Gelman, G. Jones \& X-L Meng, 
        \textit{Handbook of Markov Chain Monte Carlo} 
        (Chapman \& Hall/CRC Press)

\bibitem[Peebles (1980)]{Pee80} 
         Peebles P.J.E. 1980,
         \textit{The large-scale structure of the universe} (Princeton Univ. Press)

\bibitem[Pogosyan et al. (1998)]{pogosyan1998} 
        {Pogosyan} D., {Bond} J.R., {Kofman} L. \& {Wadsley} J. 1998,
        in S. Colombi, Y. Mellier eds., 
        \textit{Wide Field Surveys in Cosmology, 14th IAP meeting) (Editions Frontieres}, p. 61

\bibitem[Springel et al. (2005)]{millenium05}	
  	 Springel, V. et al. 2005,
	 \textit{Nature}, ~435, 629

\bibitem[Tegmark et al. (2004)]{TSB04} 
         Tegmark M., SDSS collaboration 2004
         \textit{Ap.J.}, 606, 702

\bibitem[van de Weygaert \& Bertschinger (1996)]{weyedb96}
	van de Weygaert R. \& Bertschinger E. 1996, 
	\textit{MNRAS}, ~281, 84

\bibitem[van de Weygaert \& Bond (2008)]{weybond2008}
	van de Weygaert R. \& Bond J.R. 2008, 
	\textit{Clusters and the Theory of the Cosmic Web}, 
        in M. Plionis, O. L\'opez-Cruz \& D. Hughes eds., 
        \textit{A Pan-Chromatic View of Clusters of Galaxies and the Large-Scale Structure}, 
        LNP 740 (Springer), p. 335

\bibitem[Zaroubi et al. (1995)]{zaroubi95}
        Zaroubi S., Hoffman Y., Fisher K.~B. \& Lahav O. 1995,
        \textit{ApJ}, 449, 446

\bibitem[Zeldovich (1970)]{zeldovich70}
        Zeldovich, Ya. B. 1970,
        \textit{AA}, 5, 84

\end{thebibliography}
\end{document}